\documentclass[acus]{pac99}
\usepackage{epsfig}
\newcommand{\bit}{\begin{Itemize}}
\newcommand{\eit}{\end{Itemize}}
\begin{document}

\thispagestyle{empty}

\onecolumn

\begin{flushright}
{\large
SLAC--PUB--8096\\
March 1999\\}
\end{flushright}

\vspace{.8cm}

\begin{center}

{\LARGE\bf
The Next Linear Collider Extraction Line Design\footnote
{\normalsize{Work supported by
Department of Energy contract  DE--AC03--76SF00515.}}}

\vspace{1cm}

\large{
Y.~Nosochkov,
T.~O.~Raubenheimer,
K.~Thompson and
M.~Woods \\
Stanford Linear Accelerator Center, Stanford University,
Stanford, CA  94309\\
}

\end{center}

\vfill

\begin{center}
{\LARGE\bf
Abstract }
\end{center}

\begin{quote}
\large{
The two main functions of the NLC extraction line include:
1) transmission of the outgoing disrupted beam and secondary particles to
the dump with minimal losses; and
2) beam diagnostics and control. In this report, we describe the extraction
line optics, present the results of tracking studies, and discuss
the extraction line instrumentation.
}
\end{quote}

\vfill

\begin{center}
\large{
{\it Presented at the 1999 IEEE Particle Accelerator Conference (PAC99)\\
New York City, New York, March 29 -- April 2, 1999} \\
}
\end{center}

\newpage

\pagenumbering{arabic}
\pagestyle{plain}

\twocolumn

\title{
THE NEXT LINEAR COLLIDER EXTRACTION LINE DESIGN\thanks
{Work supported by the Department of Energy Contract
DE-AC03-76SF00515.}
}

\author{
\underbar{Y.~Nosochkov\thanks{E-mail: yuri@slac.stanford.edu.}},
T.~O.~Raubenheimer,
K.~Thompson and
M.~Woods \\
Stanford Linear Accelerator Center, Stanford University, Stanford, CA 94309
}

\maketitle

\begin{abstract}
The two main functions of the NLC extraction line include: 
1) transmission of the outgoing disrupted beam and secondary particles to 
the dump with minimal losses; and 
2) beam diagnostics and control. In this report, we describe the extraction 
line optics, present the results of tracking studies, and discuss 
the extraction line instrumentation.
\end{abstract}

\vspace{-1mm}
\section{INTRODUCTION}
The power of the Next Linear Collider (NLC) \cite{y:dl:param} beams at
1 TeV (cms energy) can be as high as 10 MW and has to be safely disposed
after the interaction point (IP).  For the NLC beam parameters at the
IP, a significant disruption \cite{y:dl:holleb} of the beam
distribution occurs due to the beam-beam interaction, notably an
increase in the beam angular divergence and energy spread. In
addition, the beam collisions generate a significant amount of
beamstrahlung photons, low energy $e^+e^-$ pairs and other secondary
particles; the number of the beamstrahlung photons from the IP is 
comparable to that of the primary beam particles. The need to transport 
the photon power to the dump places constraints on the extraction line
design. Following earlier studies \cite{y:dl:spencer}, the current
design is based on a shared dump for the primary leptons and
photons. To minimize beam losses due to large energy spread in the
disrupted beam, it is critical to design optics with large chromatic
bandwidth. Additional constraints are imposed by planned diagnostics
after IP.

Several scenarios are currently under study for NLC beam parameters 
\cite{y:dl:param}. 
In this paper, we present the results for one scenario
which gives the largest energy spread and beam loss for the disrupted
beam. This set of beam parameters includes: 1046 GeV cms energy,
120 Hz repetition rate, 95 bunches per RF pulse and $0.75\cdot10^{10}$
bunch charge (see also Table~1). Since the colliding beam parameters are 
identical at IP, the extraction line design described below is applicable 
to both beams.

\vspace{-1mm}
\section{LATTICE}

The primary requirement for the NLC extraction line is to transport 
the outgoing beams to the dump with minimal losses and provide conditions 
for beam diagnostics. The main optics includes: 1) a set of quadrupoles 
after IP to focus the outgoing lepton beam; 2) a horizontal
chicane and secondary IP in the middle of chicane for beam measurements; 
and 3) a second quadrupole set at the end of the line to make a
parallel beam at the dump. The strength of the first quadrupole system
is defined by the point-to-point transformation from IP to secondary IP 
($R_{12}$=$R_{34}$=0), and the second set of quadrupoles provides a
point-to-parallel transformation to the dump ($R_{22}$=$R_{44}$=0).
The optics calculations were made using MAD code \cite{y:dl:mad}.

\begin{figure}[tb]
\includegraphics{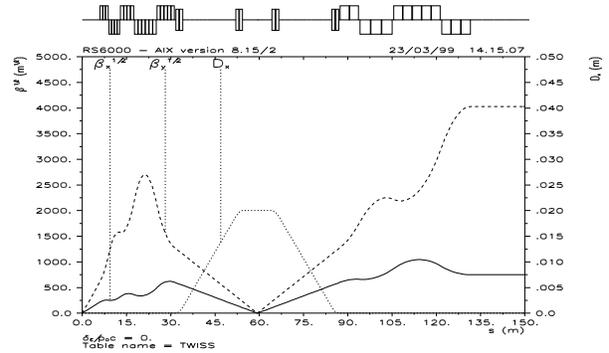}
\vspace{49mm}
\caption{Lattice functions in the extraction line.}
\label{y:dl:beta}
\vspace{-4mm}
\end{figure}

The lattice functions of the extraction line are shown in 
Fig.~\ref{y:dl:beta}.
Since the beam size at IP is much smaller in the vertical plane,
the first quadrupole after IP focuses the beam horizontally.
This minimizes the overall beam size in the extraction line.
For realistic magnet design, we use the quadrupole pole tip field 
$\leq$12 kG at 1 TeV (cms energy) and 8.2 kG field in the bends. 
The total length of the beam line is about 150 m.

The beam line optics is constrained by the  following parameters
and requirements:

\vspace{1mm}
\noindent
- Crossing angle and positions of the final focus quads, \\
- Disrupted beam parameters at IP, \\
- Angular divergence of the beamstrahlung photons, \\
- Shared dump for a primary beam and photons, \\
- Secondary IP and chicane for beam diagnostics.

\vspace{-2mm}
\subsection{IP Constraints}

The NLC beams cross at 20 mrad horizontal angle, and the nearest final focus 
quadrupoles are placed 2 m before the IP. To minimize geometric interference
between the final focus and extraction line magnets, the latter should be
placed as far as possible from the IP. However, a long free space after 
IP increases the beam size, apertures and length of the extraction line 
quadrupoles. In this design, we place the first extraction line quadrupole
6 m after IP. This clears the first three final focus 
quadrupoles.

\vspace{-2mm}
\subsection{Disrupted Beam Parameters}

\begin{figure}[tb]
\includegraphics{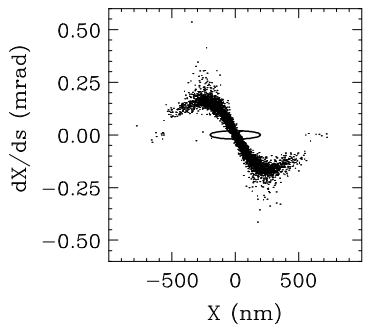}
\vspace{43mm}
\caption{Horizontal phase space at IP: dots - disrupted beam,
ellipse - 1$\sigma$ undisrupted beam.}
\label{y:dl:xphsp}
\vspace{-4mm}
\end{figure}

The strong beam-beam interactions change the beam parameters 
at IP. Notably, a significant increase occurs in the beam energy 
spread and horizontal angular divergence. The horizontal
phase space for the disrupted primary beam at IP is shown in 
Fig.~\ref{y:dl:xphsp}.
The beam distribution was calculated using GUINEA PIG beam-beam
simulation code \cite{y:dl:schulte}. The undisrupted 1$\sigma$ phase 
ellipse is shown for comparison in Fig.~\ref{y:dl:xphsp}. The nominal
and disrupted beam parameters are given in Table~1.

\begin{table}[tb]
\begin{center}
\caption{Beam parameters at IP.}
\vspace{-0mm}
\medskip
\begin{tabular}{lcc}
\hline
\textbf{Beam parameter} & \textbf{Undisrupted} & \textbf{Disrupted} \\
 & \boldmath{$x/y$} & \boldmath{$x/y$} \\
\hline
Emit. (m$\cdot$rad) [$10^{-13}$] & 39 / 0.59 & 120 / 1.02 \\
rms size (nm) & 198 / 2.7 & 198 / 3.2 \\
rms divergence ($\mu$rad) & 20 / 22 & 125 / 33 \\
$\beta^{*}$ (mm) & 10.0 / 0.125 & 3.259 / 0.103 \\
$\alpha^{*}$ & 0 / 0 & 1.805 / 0.306 \\
\hline
\end{tabular}
\label{y:dl:param}
\end{center}
\vspace{-7mm}
\end{table}

\begin{figure}[t]
\includegraphics{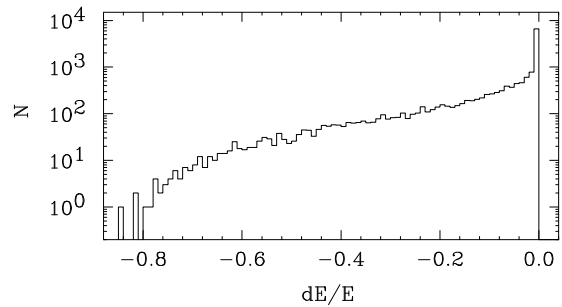}
\vspace{40mm}
\caption{Energy distribution for disrupted beam.}
\label{y:dl:espread}
\vspace{-4mm}
\end{figure}

The energy distribution for the disrupted beam is shown in 
Fig.~\ref{y:dl:espread}.  
The low energy tail extends to $\delta\sim -90$\% 
($\delta$=$\Delta p/p$), and up to 1\% of the beam power ($\sim$100 kW) 
is carried by the particles with $\delta$$<$-50\%. To minimize losses
in this energy range, the optics requires a huge chromatic 
bandwidth and large magnet apertures. The methods used to improve
the chromatic transmission are discussed below.
 
\vspace{-2mm}
\subsection{Chromatic Bandwidth}

To satisfy optics conditions for the nominal energy, the use of quadrupole 
doublets in the beginning and end of the extraction line is sufficient.
However, the strong doublets significantly overfocus the particles in the
low energy range and lead to beam losses. To reduce the overfocusing, 
we replaced the doublets by 5 alternating gradient quadrupoles 
in the beginning and 4 quadrupoles at the end of the beam line
(see Fig.~\ref{y:dl:beta}).
Since the net focusing for the nominal energy has to remain the same,
the strengths of individual quadrupoles are reduced. As a result, the
low energy particles experience less focusing in each quadrupole
and oscillate through the alternating gradient system with less
overfocusing. The strengths of individual quads were optimized by 
minimizing the low energy betatron amplitudes. The limitations of the 
described multi-quad system are the increased length of the focusing system 
and the large beam size and magnet apertures.

A simplified explanation of the multi-quad bandwidth can be made using
analogy with a FODO system of $n$ identical cells with fixed total phase
advance $\mu$. In such a system, the range of low energies satisfying
stability conditions increases with $n$ proportional to
1-$\sin(\mu/2n)$.

\vspace{-2mm}
\subsection{Beamstrahlung Photons}

Bending of the particle orbits due to beam-beam forces at IP results
in radiation and significant flux of beamstrahlung photons from IP.
The GUINEA PIG simulation shows that the rms angular spread of the
photons is on the order of $\pm$100 $\mu$rad in the horizontal plane and
a factor of 2 smaller in the vertical plane. In this design, the
primary beam and the photons are transported to one shared dump. 
For beam diagnostics in the extraction line, it is desirable to avoid
any material in the path of the beamstrahlung photons.
Therefore, large apertures of the magnets and beam pipe have to be used
to include the photon flux. For this design, we assumed the maximum
photon beam divergence of $\pm$1 mrad and $\pm$0.5 mrad in the horizontal
and vertical planes, respectively.

\vspace{-2mm}
\subsection{Chicane}

The horizontal chicane allows to separate
the outgoing electron and photon beams for measurements. In this design,
the chicane is made of 4 pairs of bends which produce a closed bump with
2 cm of horizontal displacement and dispersion. This dispersion is 
sufficient to measure the energy spread of the undisrupted beam at the 
secondary IP. Since there are no quadrupoles between bends, the orbit bump 
is closed for all energies. The maximum displacement for the low energy 
particles increases with $1/E$.

\vspace{-2mm}
\subsection{Magnet Apertures}

\begin{figure}[tb]
\includegraphics{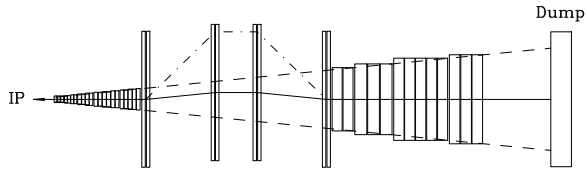}
\vspace{24mm}
\caption{Aperture constraints: solid line - nominal beam;
dash: $\pm$1 mrad photon $x$-angle; 
dash-dot: $\delta$=-90\% $x$-orbit.}
\label{y:dl:aper}
\vspace{-4mm}
\end{figure}

Large physical aperture is required for maximum transmission
of the beam to the dump. We determined the extraction line aperture by:
1) 10$\sigma$ beam size; 2) low energy horizontal excursions in 
the chicane; and 3) the maximum photon flux size. Schematically, this is 
shown in Fig.~\ref{y:dl:aper}. Outside the chicane region the aperture 
is dominated by the $\pm$1 mrad horizontal angle of the photon flux. The 
quadrupole apertures vary from $\pm$1 cm for the first quadrupole after 
IP to $\pm$13 cm near the dump. 

To minimize beam 
losses in the chicane region, we increased apertures to include low
energy orbits up to $\delta=-90$\%. With the beam size included,
the maximum horizontal aperture in the chicane is $\pm$20 cm. Since 
the orbit excursions in the chicane
occur in the horizontal plane, a smaller vertical aperture can be used
in the bends. The tracking simulations showed that $\pm$50 mm vertical gap
in the bends is sufficient to minimize beam losses and include the
$\pm$0.5 mrad vertical photon angle.

\vspace{-1mm}
\section{DIAGNOSTICS}

Beam line diagnostics fall into three categories:  1) standard diagnostics
(BPMs, toroids and ion chambers) to facilitate cleanly transporting the beam 
to the beam dump; 2) luminosity diagnostics to measure and optimize
the luminosity; and 3) physics diagnostics to measure
the beam polarization, energy, and energy spread.  

The luminosity diagnostics
will include BPMs with 1 $\mu$m resolution for measurements of deflection
angles, as well as detectors to monitor low energy particles produced at the
IP from radiative Bhabha and pair production processes.  The physics 
diagnostics will include a Compton polarimeter, an energy spectrometer, and a 
wire scanner to measure energy spread.  The Compton polarimeter will collide 
a laser beam with the electron beam in the middle of the chicane, and its 
detector will analyze Compton-scattered electrons below 50 GeV after the
chicane.  A conventional wire scanner in the chicane can be used for the
energy spread measurements.  An SLC-style energy spectrometer is 
planned between the chicane and the beam dump to measure the separation of
synchrotron light due to a precisely calibrated spectrometer magnet.

\vspace{-1mm}
\section{BEAM LOSS}

The methods used to improve the beam transmission included: 1) the
use of multi-quad focusing systems for large chromatic bandwidth;
and 2) sufficiently large magnet apertures. To assure accuracy of 
the beam transport with low energy tail, we used a modified version of the 
DIMAD code \cite{y:dl:pt} which can handle chromatic terms to all orders.

With up to 10 MW of the NLC beam power, even the loss of just 0.3\% would
be equivalent of losing the whole SLC beam (30 kW). Therefore, our goal was
to reduce to a minimum the overall beam losses.
In addition, an excessive beam loss would interfere with the planned
diagnostics and experiments in the extraction line.

In tracking simulations, we used a disrupted distribution of 15,000
primary beam particles calculated with GUINEA PIG code. This distribution was 
tracked from IP to the dump and the beam losses were monitored along the 
beam line. We used round apertures for the quadrupoles and drifts, and
rectangular aperture for the bends.

\begin{figure}[tb]
\includegraphics{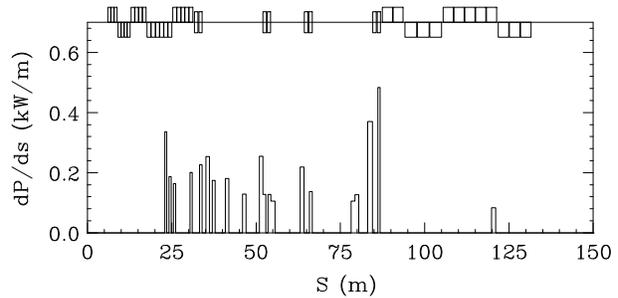}
\vspace{39mm}
\caption{Distribution of the beam power loss.}
\label{y:dl:power}
\vspace{-4mm}
\end{figure}

The most losses occur for the very low energy particles
which experience strong overfocusing in quadrupoles and
large horizontal deflections in the bends.
The calculated beam power loss along the beam line is shown in
Fig.~\ref{y:dl:power}. In this case,
all particles with $\delta$$>$-50\% and most with the lower energies 
are transported to the dump.
The total loss is 4.7 kW (0.25\% particles) and the distributed 
power loss is below 0.5 kW/m. At the dump, the rms ($x/y$) beam size is
$7.7/4.7$ mm with the tails extending to $\pm$$100/40$ mm.

The simulations included the 6 T detector solenoid 
(12 Tm after IP). Due to the crossing angle, the solenoid induces vertical
orbit distortions. The study showed that with corrected vertical
orbit after IP, the solenoid effect on the beam loss is negligible.

\vspace{-1mm}
\section{FUTURE STUDIES}

In future studies, we need to increase statistics of GUINEA PIG simulations
to obtain more accurately the distributions of electrons
in the low energy tail and beamstrahlung photons at large angles.
Tracking of these particles is needed to calculate signal to background 
ratios in the diagnostic detectors.

More details have to be included in the design of the beam line 
diagnostics, in particular the magnets for the energy spectrometer.

Methods of directing the main beam and photons to separate dumps
and possibly reducing the backgrounds and neutron back-shine from
the dump need to be investigated.

\end{document}